\documentclass[twocolumn,pra,aps,superscriptaddress]{revtex4}

\usepackage{amsthm}
\usepackage{amsmath,latexsym,amssymb,verbatim,enumerate,graphicx}
\usepackage{hyperref}
\usepackage{color}

\newtheorem{thm}{Theorem}
\newtheorem*{thm*}{Theorem}

\newtheorem*{prop*}{Proposition}
\newtheorem{lem}[thm]{Lemma}
\newtheorem*{lem*}{Lemma}

\newtheorem*{fact*}{Fact}

\newtheorem*{cor*}{Corollary}

\newcommand{\non}{\nonumber}

\newcommand{\bthm}{\begin{thm}}
\newcommand{\ethm}{\end{thm}}
\newcommand{\blem}{\begin{lem}}
\newcommand{\elem}{\end{lem}}

\newcommand{\ot}{\otimes}
\newcommand{\<}{\langle}
\renewcommand{\>}{\rangle}

\DeclareMathOperator{\rank}{rank}

\DeclareMathOperator{\poly}{poly}
\DeclareMathOperator{\tr}{tr}

\def\veps{\varepsilon}

\def\cO{{\cal O}}

\def\cL{{\cal L}}

\def\cB{{\cal B}}

\def\tpi{{\tilde \pi}}

\def\Th{ \Theta}

\begin{document}

\title{Locality of Entanglement Spectrum and Edge States from Strong Subadditivity}

\author{Kohtaro Kato}
\affiliation{Department of Physics, Graduate School of Science, The University of Tokyo, Tokyo, Japan}
\affiliation{Institute for Quantum Information and Matter  \\ California Institute of Technology, Pasadena, CA 91125, USA}

\author{Fernando G.S.L. Brand\~ao}
\affiliation{Institute for Quantum Information and Matter  \\ California Institute of Technology, Pasadena, CA 91125, USA}
\affiliation{Google Inc., Venice, CA 90291, USA}

\begin{abstract}
We consider two-dimensional states of matter satisfying an uniform area law for entanglement. We show that the topological entanglement entropy is equal to the minimum relative entropy distance from the reduced state to the set of thermal states of local models. The argument is based on strong subadditivity of quantum entropy. For states with zero topological entanglement entropy, in particular, the formula gives locality of the states at the boundary of a region as thermal states of local Hamiltonians. It also implies that the entanglement spectrum of a two-dimensional region is equal to the spectrum of a one-dimensional local thermal state on the boundary of the region.

\end{abstract}
\maketitle
\section{Introduction}
Topologically ordered phases, which appear e.g. in fractional quantum Hall systems~\cite{PhysRevB.40.7387,PhysRevB.41.9377} and in quantum spin liquids~\cite{Kitaev2003a,PhysRevB.71.045110}, are quantum phases in gapped systems which go beyond the conventional paradigm of symmetry-breaking. 
Systems in topologically ordered phases have several distinct features: topology-dependent ground state degeneracy, locally indistinguishable ground states which cannot be created by a constant-depth local circuit, and anyonic excitations. These characteristic properties are robust against local perturbations and such phases are considered as a candidate of the stage to perform fault-tolerant quantum information processing.

In the last decades, studying entanglement in quantum states has shown to be a powerful tool to characterize topologically ordered phases. 
One distinctive aspect of entanglement in ground states of gapped systems (gapped ground states) is that it satisfies an area law: the entanglement entropy scales only as the perimeter instead of the volume of a region, which is true for Haar-random states~\cite{Hayden2006}. Especially, the area law of  ground states in topologically ordered phases contain a characteristic term called the topological entanglement entropy (TEE)~\cite{PhysRevLett.96.110404, PhysRevLett.96.110405}. TEE only depends on the type of the phase and has been used as a probe of topological order~\cite{Furukawa2007,Haque2007,Isakov2011,Depenbrock2012}. 

Topological entanglement entropy has been linked to several other aspects of topological order. 
If TEE is zero, then the state can be created by a constant-depth local circuit, and thus in a topologically trivial phase ~\cite{Kitaevt13,PhysRevB.94.155125,2016arXiv160907877B}. Also, TEE upper bounds the logarithmic of the topological degeneracy of the model~\cite{PhysRevLett.111.080503}. 
Finally, TEE has also been argued to give the logarithmic of the total quantum dimension of the anyonic excitations of the system~\cite{PhysRevLett.96.110404, PhysRevLett.96.110405}.

The entanglement entropy of a region $R$ is a function of the eigenvalues of the reduced state $\rho_R$ on $R$. It is interesting to explore which information might be encoded in the whole spectrum of $\rho_R$ (i.e. all its eigenvalues). Since $\rho_R$ is positive semi-definite, we can write $\rho_R = e^{-H_R}$ for a Hermitian operator $H_R$. The operator $H_R$ is called the {\it entanglement Hamiltonian} (or modular Hamiltonian) and its eigenvalues are called the {\it entanglement spectrum}. Starting with the work of Li and Haldane \cite{PhysRevLett.101.010504}, the behavior of the entanglement spectrum of two-dimensional systems has been extensively studied.  
Based on numerical calculations~\cite{PhysRevB.83.245134,PhysRevLett.111.090501}, it was observed that for gapped systems with no topological order, one could equate the entanglement spectrum to the spectrum of a one-dimensional quasi-local Hamiltonian acting on the boundary of the region $R$. While for topologically ordered systems, a universal non-local interaction emerges. However so far it has been a challenge to give a more general argument for the locality of the entanglement spectrum, except some exact renormalization fixed-points in the tensor network formalism~\cite{Cirac17}.

A natural question is if these two aspects of entanglement in topological order are related. In this paper we  explicitly construct a quantitative relation between TEE and the entanglement Hamiltonian by showing that the TEE equals (half) the minimum relative entropy of the reduced state on annular region (which we call edge state) to the set of Gibbs states $e^{-H}$ with local Hamiltonian $H$.  
Using this result, we will give a general argument for the locality of entanglement spectrum of certain regions and its relation to the TEE.  
Our approach will be information theoretical. In particular we will derive our results from the strong subadditivity property of the von Neumann entropy and a recent strengthening thereof~\cite{Fawzi2015}. Furthermore, our result provides an information-theoretic interpretation to TEE as the number of bits of information needed to describe the non-local properties of the edge state of the system.  

\vspace{0.1 cm}
\section{ Assumption: uniform area law}  
In this work, we consider quantum systems on two-dimensional spin lattices with local dimension $d$. 
$|R|$ denotes the number of sites in region $R$ of a lattice, and $|\partial R|$ denotes its perimeter length. We will be concerned with pure states $\rho=|\psi\>\<\psi|$ on the lattice satisfying an area law: for every simply connected contractible region $R$, the von Neumann entropy $S(R)_\rho = - \tr(\rho_R \log \rho_R)$ (with $\rho_R$ the reduced density matrix of the state in region $R$) obeys
\begin{equation}\label{arealaw}
S(R)_\rho =  \alpha  |\partial R | - \gamma + c + \veps,
\end{equation}
for constants $\alpha,c, \gamma,\veps \geq 0$ ($\gamma$ is replaced by $n_R\gamma$ when $R$ has $n_R$ distinct boundaries). The constant term $\gamma$ is the topological entanglement entropy (TEE)~\cite{PhysRevLett.96.110404, PhysRevLett.96.110405}. TEE is related to the theory of anyon models via
\begin{eqnarray}
\gamma=\log\sqrt{\sum_ad_a^2}\,,
\end{eqnarray}
where $d_a\geq1$ is the quantum dimension associated to anyonic charge $a$. In topologically trivial systems, there is only the  vacuum charge ``$1$'' with $d_1=1$, and thus $\gamma=0$. 
The term $c$ gives the contribution from the corners of the region to the entanglement entropy and has the form:
\begin{equation}
c=\beta\sum_i\nu(\theta_i)\,,
\end{equation}
for a constant $\beta$ and function $\nu$. The sum is over all corners of the region, each with angle $\theta_i$. The last term $\veps$ stands for sub-leading terms in ${ o}(1)$ which go to zero when the minimum length of the region grows. 

In particular, throughout this work we require that the area law is \textit{uniform}, in the sense that the parameters $\alpha$ is independent of the choice of the region $R$. We further require $\varepsilon = \exp(- l / \xi)$, with $l$ the minimum length of the region and $\xi$ a constant (which can however be much larger than the correlation length of the system), which we expect to hold for generic gapped ground states; see Appendix~\ref{ap4}. Note that our result still holds if $\veps$ decays polynomially but sufficiently fast. 

\vspace{0.1 cm}
\section{Definition of edge states and the main formula} 
\begin{figure}[htbp]  
\begin{center}
\hspace{0mm}
\includegraphics[width=6.0cm]{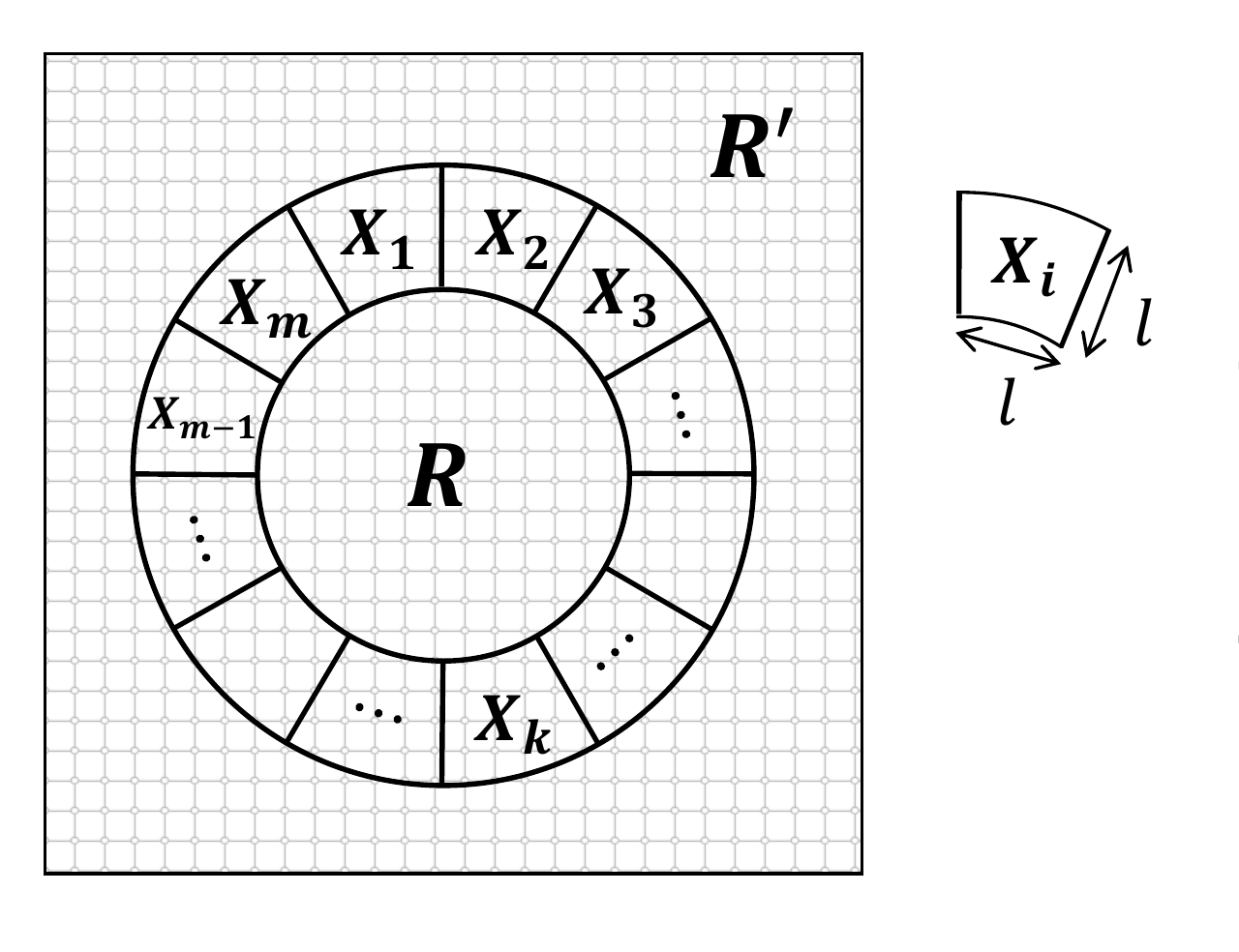}
\vspace{-5mm}
\end{center}
\caption{Region $R$, its boundary region $X$ and the complement $R'$. The size of each region $X_i$ is specified by $l$. }
\label{ring}
\end{figure}

Consider a region $R$ with a boundary region $X$ as in Fig.~\ref{ring}. $X$ is composed by $m$ regions $X_i$, each with length scale $l$. We can regard $X=X_1X_2...X_m$ as a one-dimensional spin system with $X_i$ has  local dimension $d^{|X_i|}$. We say $\rho_X$, the reduced density matrix of $|\psi\>$ on the boundary $X$, is the {\it edge state} of the region $R$. We could take $R$ as large as the whole lattice, in which case $X$ would indeed be the physical edge of the system. However our result also holds when $R$ is a subregion of the entire lattice (in this case $X$ corresponds to the entanglement cut between $R$ and $R'$).

An important quantity in our approach is the conditional mutual information, defined for tripartite states $\rho_{ABC}$ as 
\begin{align}
I(A:C|B)_{\rho}:=S(AB)_{\rho} + S(BC)_{\rho} - S(ABC)_{\rho} - S(B)_{\rho}. \nonumber
\end{align}
It is a measure of the correlations between $A$ and $C$ conditioned on the information in $B$. The strong subadditivity inequality of von Neumann entropy~\cite{SSA73} reads $I(A:C|B)_{\rho} \geq 0$. As observed in \cite{PhysRevLett.96.110405}, the uniform area law~\eqref{arealaw} implies that for every (connected) triple $ABC$ with $A$ and $C$ disconnected, conditional mutual information has a dichotomy of values:  $I(A:C|B) \approx 0$ if $ABC$ is topologically trivial, while $I(A:C|B) \approx 2 \gamma$ if it is topologically non-trivial annulus. 

The main formula of this paper is a new characterization of TEE in terms of the relative entropy distance between the edge state and the set of thermal states of local models. 
Define the set of Gibbs states of short-range Hamiltonians with interaction strength $K$ as 
\begin{equation}
E^K_{nn}:=\left\{e^{-H}\left|\; H=\sum_ih_{X_iX_{i+1}},\; \|h_{X_iX_{i+1}}\|\leq K\right.\right\}\,.
\end{equation}
Note that here we include the normalization factor in the Hamiltonian so that $\tr(e^{-H})=1$. 
Then, we can show that 
\begin{equation} \label{TEErelent}
\gamma \approx \frac{1}{2} \min_{e^{-H} \in E^K_{nn}}  S \left( \rho_X  \left\Vert  e^{-H}  \right. \right)\,
\end{equation}
for $K=\Th(N)$, where $\approx$ means the equality holds up to $\cO(e^{-\Th(l)})$ if we choose $l=\log|X|$~\footnote{$x$ is in $\Th(l)$ if there exists $l_0>0$ such that there exist two constants $c,C>0$ satisfying $cl\leq x\leq Cl$ for any $l\geq l_0$.}. For $\veps=0$ in which we have exact equality, the formula was proven before by one of us in Ref.~\cite{PhysRevA.93.022317}. Each term of $H$ in $E_{nn}^K$ acts on at most $O(\log(|X|))$ sites and thus $H$ is a (quasi-)local Hamitonian. 
Note that numerical results of Refs. \cite{PhysRevB.83.245134,PhysRevLett.111.090501} suggest that one might be able to improve Eq.~\eqref{TEErelent} to have Hamiltonians with exponentially-decaying interactions with locality independent of system size. 
Which entanglement Hamiltonian achieves the minimum in Eq.~\eqref{TEErelent}? Although we do not know the answer, 
\begin{equation}
H_X:=-\sum_i\left(\ln\rho_{X_iX_{i+1}}-\ln\rho_{X_i}\right)
\end{equation}
could be a natural guess. Actually, one can show that unnormalized Gibbs state $e^{-H_X}$ has distance close to $2\gamma$ (see Appendix~\ref{ap1} and Ref.~\cite{PhysRevB.87.155120}). Notably, this (possibly unbounded) local Hamiltonian is calculable only from local reduced states.

Equation (\ref{TEErelent}) also provides an information-theoretic interpretation for TEE. Let us recall a result of Ref.~\cite{anshu2017quantum}. Consider two parties, Alice and Bob. Alice (Bob) has a classical description of the density matrix $\rho$ ($\sigma$). They also share unlimited entanglement. Then Alice can send $S(\rho \Vert \sigma)/2$ qubits to Bob such, after a decoding operation by Bob, he has a quantum state which is close to $\rho$ (the error goes to zero in the asymptotic regime, where one consider the protocol applied to $\rho^{\otimes n}/\sigma^{\otimes n}$ for very large $n$). Moreover, there is no protocol with a lower rate \cite{anshu2017quantum}. Therefore the relative entropy $S(\rho \Vert \sigma)$ has the interpretation of the number of qubits which are contained in $\rho$ in addition to the information contained in $\sigma$. Applied to our setting, Eq. (\ref{TEErelent}) can then be interpreted as saying that TEE gives the number of qubits which are contained in the edge state in addition to any local model; it counts the number of topological qubits of the model.

\vspace{0.1 cm}
\section { Entanglement spectrum on a cylinder} 
For a pure bipartite state $|\psi\>_{AB}$, consider the Schmidt decomposition: 
\begin{equation}\label{Schdeco}
|\psi\>_{AB}=\sum_i\sqrt{\lambda_i}|i\>_A|i\>_B\,,
\end{equation}
where $\{|i\>_A\}$ and $\{|i\>_B\}$ are orthonormal vectors of systems $A$ and $B$. The coefficients $\lambda_i$ satisfying $\lambda_i>0$ and $\sum_i\lambda_i=1$ are called the Schmidt coefficients.  
The entanglement spectrum of $\rho_R$ is defined by $\{-\log\lambda_i\}_i$. Note that Eq.~\eqref{Schdeco} shows that the entanglement spectrum on a subsystem $R$ always matches to the spectrum on the complement. 

\begin{figure}[htbp]  
\begin{center}
\hspace{-3mm}
\includegraphics[width=6.0cm]{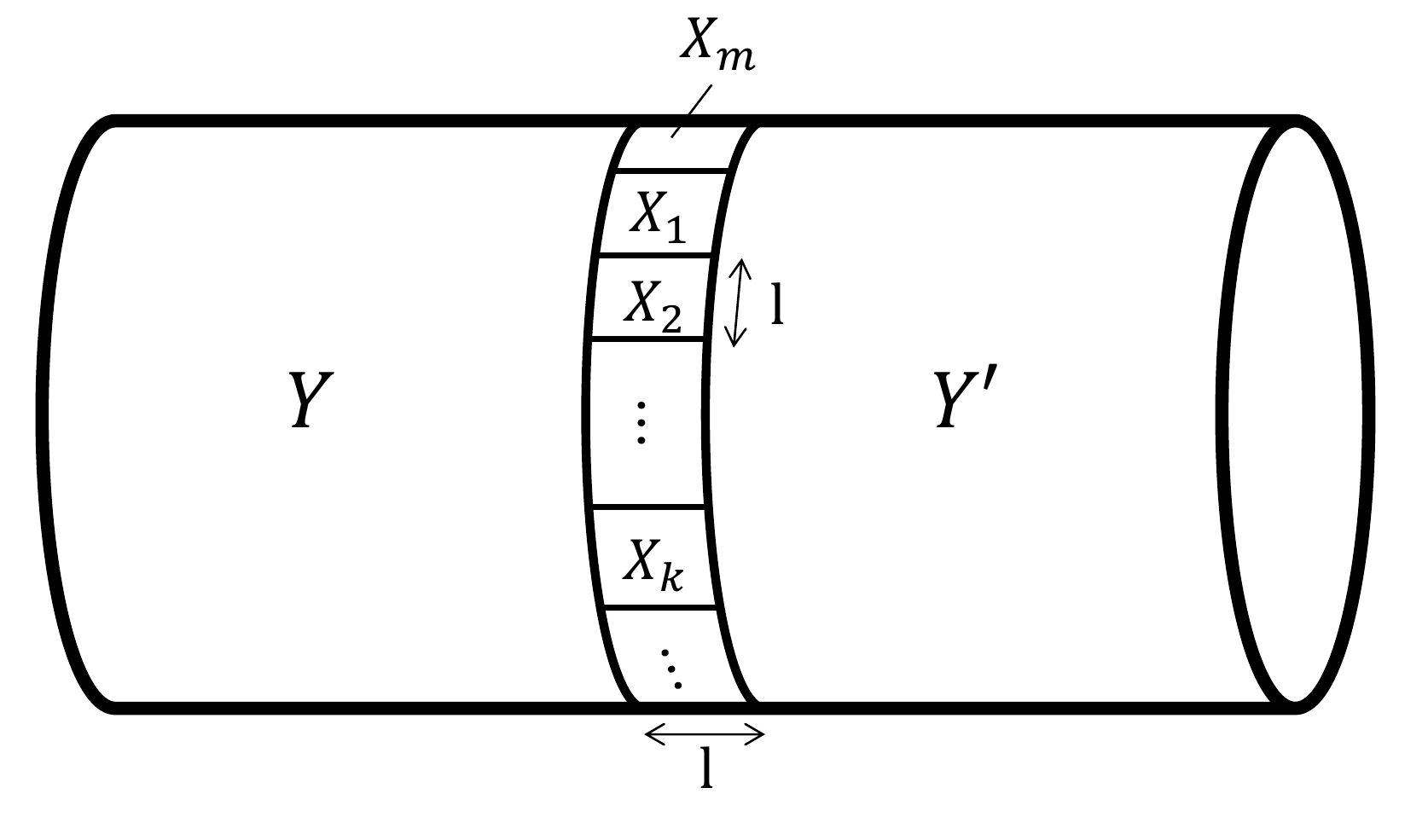}
\vspace{-5mm}
\end{center}
\caption{We consider a system on a 2D cylinder.  We divide it into three regions $Y$, $X$ and $Y'$ so that $X$ can be viewed as a 1D \lq\lq{}boundary\rq\rq{} of $Y$ as in Fig.~\ref{ring}. }
\label{cyl}
\vspace{-2.5mm}
\end{figure}

Let us now turn to the application of Eq.~\eqref{TEErelent} to analyze the structure of the entanglement spectrum of the system.  
For concreteness, we consider the entanglement spectrum of a system defined on a cylinder.  
Consider a ground state of a system as depicted in Fig.~\ref{cyl}. 
Then the spectrum (of the reduced state) on region $YY'$ is the same as the spectrum on region $X$. 
Let us assume that the system has reflection symmetry, so that $\rho_Y=\rho_{Y'}$.  
For a ground state in a topologically trivial phase satisfying Eq.~\eqref{arealaw}, we have $I(Y:Y')\approx0$ which implies $\rho_{YY'}\approx\rho_Y^{\otimes 2}$ (this is followed by the fact that the ground state is approximately generated by a constant-depth circuit).  Indeed, Pinsker's inequality reads
\begin{equation}
I(Y : Y') \geq   \frac{1}{2}  \Vert \rho_{YY'} - \rho_Y \otimes \rho_{Y'} \Vert_1^2,
\end{equation}
with $\Vert \rho_{YY'}- \rho_Y \otimes \rho_{Y'} \Vert_1$ the trace-norm distance between $\rho_{YY'}$ and the product of its reductions $\rho_Y \otimes \rho_{Y'}$. 

We denote the entanglement Hamiltonian of $\rho_Y^{\otimes 2}$, which we call the double of $H_{\rho_Y}$, by $H^{(2)}_{\rho_Y}=H_{\rho_Y}\ot I+I\ot H_{\rho_Y}$ (where $I$ is the identity operator). 
We also introduce a cut-off $\Lambda$ on the spectrum of operators by
\begin{equation}
\lambda^\Lambda(A):=\left\{\lambda\in\lambda(A)\left|\lambda\leq\log\Lambda\right.\right\}\,.
\end{equation}
Then, the result on the locality of edge states~\eqref{TEErelent} implies that when $\gamma=0$, there exists a 1D nearest-neighbor Hamiltonian $H_X=\sum_ih_{X_iX_{i+1}}$ on $X=X_1...X_m$, such that for any $\Lambda>0$, 
\begin{equation}\label{thm1}
\left\|\lambda^\Lambda\left(H_{\rho_Y}^{(2)}\right)-\lambda^\Lambda(H_X)\right\|_1\leq \Lambda e^{-\Th(l)}\,
\end{equation}
(the proof is given in Appendix~\ref{ap2}).  
The upper bound decays exponentially in $l$ if we choose $\Lambda=\poly(l)$.  
Note that there exists a unique ground states of gapped models with $I(A:C|B)>2\gamma=0$ for a certain choice of region $X=ABC$~\cite{BravyiCEX,PhysRevB.94.075151}. $H_X$ turns out to be non-local in this exotic example not satisfying our assumption. However, we can recover $I(A:C|B)\approx0$ by slightly changing the shape of $X$ for these counterexamples. 

Eq.~\eqref{TEErelent} also implies that there exists an isometry $V$ from $Y^{\ot2}$ to $X$ such that 
\begin{equation}
V\rho_{Y}^{\otimes 2}V^\dagger=e^{-H_{\rho_X}}\approx e^{-\sum_ih_{X_iX_{i+1}}}\,
\end{equation}
(here $\approx$ means both sides are exponentially close  with respect to $l$ in the relative entropy/the trace distance). 
When $\rho_Y$ has a symmetry under some unitary $U$, $U\rho_YU^\dagger=\rho_Y$, 
the edge state have a corresponding symmetry 
\begin{equation}
U'\left(e^{-H_{\rho_X}}\right)U'^\dagger=e^{-U'H_{\rho_X}U'^\dagger}=e^{-H_{\rho_X}}
\end{equation}
for any $U'$ such that $U'V=VU$. 

In topologically ordered phases, one can naturally expect that the entanglement Hamiltonian $H_{\rho_X}$ should be non-local due to a non-zero TEE. 
However, we have to be careful since it is known that the sub-leading term in Eq.~\eqref{arealaw} for a non-contractible region (like $X$) not only depends on the type of the phase, but also depends on the choice of the ground state~\cite{PhysRevB.85.235151,PhysRevB.94.075126}. 
For this reason, $I(Y:Y')\approx0$ does not hold for general ground states, and thus the previous argument should be suitably modified. 
Let us assume that there always exists a special orthonormal basis of the ground subspace for a gapped system such that $I(Y:Y')\approx0$ holds for each basis element. 
This assumption is reasonable if the ground subspace is spanned by minimally-entangled states~\cite{PhysRevB.85.235151}  $\{|\psi_a\>\}_a$, which have a definite anyonic flux threading through the cylinder labeled by a finite set $\cL=\{a\}$. 
For such states, we expect the modified area law
\begin{equation}\label{modarealaw}
S(R)_\rho =  \alpha  |\partial R | - 2\gamma+\log d_a + c + \varepsilon\,,
\end{equation}
where $d_a$ is the quantum dimension of the anyon flux $a$, to hold for any non-contractible subregion $R$ on the cylinder, as $X$ in Fig.~\ref{cyl}.  
Then, there exists a 1D Hamiltonian $H^a_X$ on $X=X_1...X_m$ for each $a\in\cL$, such that for any $\Lambda>0$, 
\begin{equation}\label{eq:thm2}
\left\|\lambda^\Lambda\left(H_{\rho^a_Y}^{(2)}\right)-\lambda^\Lambda(H^a_X)\right\|_1\leq \Lambda e^{-\Th(l)}\,,
\end{equation}
with $\rho_Y^a=\tr_{XY'}|\psi_a\>\<\psi_a|$. Here we again assume the reflection symmetry. Importantly, here $H^a_X$ contains non-local interactions in contrast to the case of $\gamma=0$.

A general ground state $|\psi\>=\sum_{a\in{\cal L}}\sqrt{p_a}|\psi_a\>$ is a superposition of states with different fluxes. 
Each anyonic flux $a$ can be measured by a projective measurement 
acting on $YY'$, and therefore the reduced states on $YY'$ with different fixed anyonic flux are orthogonal. 
Hence, we have a direct sum decomposition of the reduced state:
\begin{equation}
\rho_{YY'}=\bigoplus_{a\in{\cal L}}p_a\rho^a_{YY'}\,.
\end{equation}
Using the reflection symmetry and $I(Y:Y')\approx0$ for each $a$, we have 
\begin{equation}\label{directsumrdm}
\rho_{YY'}\approx \bigoplus_{a\in{\cal L}}p_a\rho^{a\otimes2}_{Y}\,.
\end{equation}
As in the case of the trivial phase,  there exists an isometry $V$ from $YY'$ to $X$ such that
\begin{equation}
V\rho_{YY'}V^\dagger\approx\sum_ap_ae^{-\sum_ih^a_{X_iX_{i+1}}-h_X^a}\,,
\end{equation}
where $h_X^a$ acts on $X$ non-locally. We expect that each $h^a_X$ represent a topological constraint and is dominated by $m$-body interactions (as we discuss in Appendix~\ref{ap1}). Indeed this has been observed before for some exactly solvable models~\cite{PhysRevB.83.245134,PhysRevLett.111.090501,PhysRevA.93.022317}. 

We have shown that the double of the entanglement spectrum is approximately equivalent to the spectrum of the 1D edge  state, which is local if TEE is zero. We now want to argue that under a few more assumptions, the same property also holds for the single entanglement spectrum. 

\begin{figure}[htbp]  
\begin{center}
\vspace{-3mm}
\includegraphics[width=6.0cm]{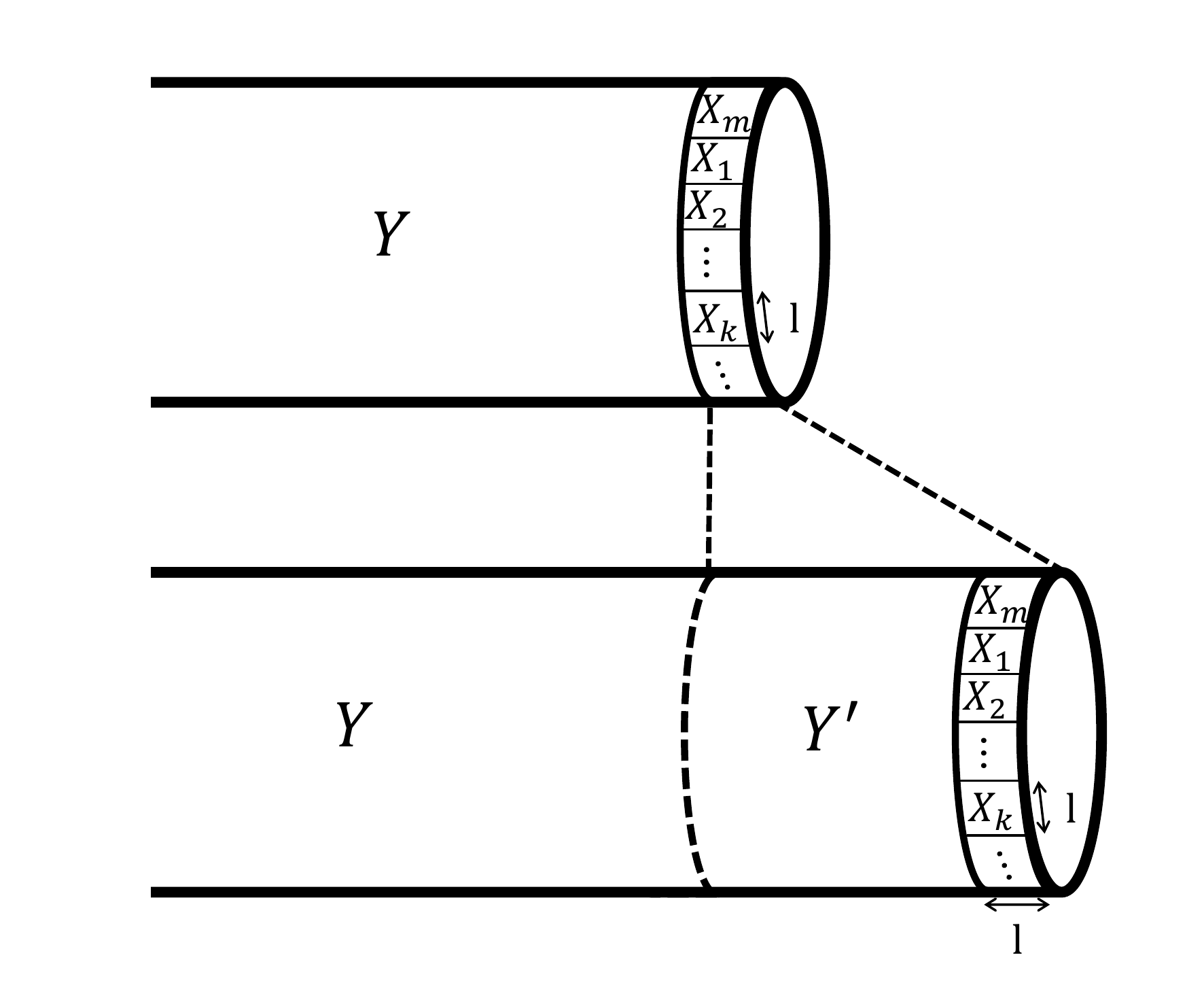}
\vspace{-5mm}
\end{center}
\caption{(Up) We choose $X$ as the region around the physical boundary (the right edge). The entanglement spectrum on $Y$ is the same as that of $X$. 
(Down) In some cases, the reduced state on $Y$ is almost independent of the length of the opposite side. 
Then the entanglement spectrum of $Y$ is equivalent to the spectrum of $X$ which is an edge of another cylinder with shorter length. }
\label{cyl2}
\end{figure}

Let us first consider a ground state on a cylinder with a boundary (or boundaries) as in the upper part of Fig.~\ref{cyl2}. 
Here we choose $X$ as a region around the physical boundary. 
The entanglement spectrum of $Y$ is equivalent to that of $X$ since the state on $XY$ is pure. 
The edge state  on $X$ depends on how we choose interaction terms around the boundary, but we can still apply Eq.~\eqref{TEErelent} if the edge state satisfies the area law of Eq.~\eqref{arealaw}. For instance, the toric code with a smooth boundary~\cite{Kitaev2012} satisfies the assumption.  
 
For more general situations, let us turn back to a ground state $|\psi\>$ of a system defined as in Fig.~\ref{cyl}.  
Remember that $\rho_{YY'}\approx\rho_Y\ot\rho_{Y'}$ if $|\psi\>$ satisfies $I(Y:Y')\approx0$. 
Consider a purification $|\psi^L\>_{YX_1}\ot|\psi^R\>_{Y'X_2}$ of $\rho_Y\ot\rho_{Y'}$ on some ancillary system $X_1X_2$ satisfying $\psi^L_Y=\rho_Y$ and $\psi^R_{Y'}=\rho_{Y'}$. By Uhlmann's theorem~\cite{UHLMANN1976273}, there exists a unitary $U_X$ from $X$ to systems $X_1$ and $X_2$ such that
\begin{equation}\label{Uhlmann}
U_X|\psi\>_{YXY'}\approx|\psi^L\>_{YX_1}\otimes|\psi^R\>_{X_2Y'}\,, 
\end{equation}
We can choose $|X_1|\sim\cO(|\partial Y|)$  {and interpret $YX_1$ as a new cylinder} if $\rho_Y$ can be well-approximated by a low-rank state ${\tilde\rho}_Y$ with $\rank({\tilde \rho}_Y)=2^{\cO(|\partial Y|)}$. 
This approximation has been shown to be possible for any ground state satisfying the area law~\cite{WE18} (not necessarily to be uniform), while the error term only decreases $\cO(\frac{1}{l})$. 
The new edge state $\psi^L_{X_1}$ on $X_1$ has almost the same spectrum as $\rho_Y$ and we can use the previous argument discussed above. 
Furthermore, the validity of the approximation is invariant under any constant-depth local circuit, since such a circuit can only add constant (of the axial length) to the rank of reduced state on $Y$. 
Therefore, all ground states in the topologically trivial phase satisfy the condition, since they can be created from product state by such circuits.

Another example is a family of gapped ground states which can be described by Matrix Product States (MPS)~\cite{MPSo08} defined in the axial direction. 
Suppose a (unnormalized) state $|\psi_{N}\>$ is defined on a cylinder with the axial length $N$ and the radius $r$. We obtain a 1D system by cutting the cylinder into several slices and then regarding one slice as one large subsystem. 
Suppose that  $|\psi_{ N}\>$ can be written as
\begin{align}
|\psi_{N}\>=\sum_{i_1,...,i_N}\left(L\right|A^{i_1}\ldots A^{i_N}\left|R\right)|i_1i_2\ldots i_N\>\,,
\end{align}
where the indices $\{i_j\}$ is associated with the  $j$th slice (column) of the cylinder, and 
$\{A^i\}_i$ are $D\times D$ matrices with a bond dimension $D\sim 2^r$. $|L)$ and $|R)$ are $D$-dimensional vector representing the boundary condition (we used ``)'' to distinguish them from vectors in physical systems). 
Choose the first $m$ slices as subsystem $Y$. Then, one can show that in generic case the reduced density matrix on $Y$ is almost independent of $N$ for sufficiently large $N$ (More details are in Appendix~\ref{ap3}). 
Therefore,  the spectrum on $Y$ is approximately equivalent to the spectrum of the edge state defined for some fixed cylinder (Fig.~\ref{cyl2}).

\section{Discussion}  In this work we have gave a new formula for TEE, connecting it to the locality of edge states. In particular, we showed that if TEE is zero, the entanglement Hamiltonian of the 1D edge state is approximately a short-range Hamiltonian, while it is a non-local Hamiltonian if the ground state have non-zero TEE. 
We then applied this result to the entanglement spectrum defined on a half of a cylinder, and derived that the double of the spectrum matches the spectrum of a 1D Hamiltonian (which is local if TEE is zero). 
We also have shown that the same results hold for the single entanglement spectrum under additional but physically reasonable assumptions. 
Our techniques only rely on the property of ground states and is independent of specifics of particular models.  

A similar connection has been observed before in the PEPS formalism, where the edge state is defined for an effective boundary on virtual degrees of freedom. In our case, the edge state is defined via the reduced state on the boundary, and therefore it acts on physical degrees of freedom. 
Building an explicit connection between our framework and the PEPS formalism is an interesting open question. Another interesting direction for future research is to weaken our assumptions and extend our results for more general gapped systems. Especially,  it is unclear if we can always find a suitable isometry in Eq.~\eqref{Uhlmann} such that the edge state on the new physical boundary satisfies the area law assumption (presently we can only show it for a few explicit examples, e.g., the toric code).

\vspace{0.2 cm}

{\bf Acknowledgements.}
We thank for Burak Sahinoglu for helpful discussions. We acknowledge support from the NSF. Part of this work was done when both of us were working in the QuArC group of Microsoft Research. KK thanks Advanced Leading Graduate Course for Photon Science (ALPS) and JSPS KAKENHI Grant Number JP16J05374 for financial support. 
\appendix

\section{A Proof of Eq.~\eqref{TEErelent}}\label{ap1}
In this appendix, we prove the main formula~\eqref{TEErelent}. 
Let us first consider an entanglement Hamiltonian 
\begin{equation}\label{eq:entHamil}
H_X:=-\sum_i\left(\ln\rho_{X_iX_{i+1}}-\ln\rho_{X_i}\right)\,,
\end{equation}
where we are using the periodic boundary conditions, so $m+1$ is identified with 1. We can write 
\begin{align}  \label{mainrel22}
S(\rho_X \Vert e^{-H_X})  = \sum_{i=1}^m S(X_{i+1} | X_{i})- S(X_1\ldots X_m), 
\end{align}
with $S(X_{i+1} | X_{i}) := S(X_{i} X_{i+1}) - S(X_{i})$ the conditional entropy of $X_{i+1}$ given $X_i$ (we omit the index $\rho$). 

Note that 
\begin{align}
&S(X_1\ldots X_k) + S(X_{k+1}|X_k) \nonumber\\
&=I(X_1 \ldots X_{k-1} : X_{k+1} | X_k) +  S(X_1\ldots X_{k+1}) \label{inyerationentropy}
\end{align}
From Eq.~\eqref{arealaw}, we have 
\begin{equation}
I(X_1 \ldots X_i : X_{i+2} | X_i) \leq \varepsilon 
\end{equation}
for $i \in \{1, \ldots, m-2 \}$ and $\veps$ given by error term in Eq.~\eqref{arealaw}. Then using Eq. (\ref{inyerationentropy}) $(m-2)$ times in Eq. (\ref{mainrel22}), 
\begin{align}
&S(\rho_X \Vert e^{-H_X}) \nonumber\\ 
&\approx_{(m-2)\varepsilon} S(X_1 \ldots X_{m-1}) + S(X_{m}|X_{m-1}) +  S(X_{1}|X_{m})  \nonumber  \\ 
&\qquad- S(X_1) - S(X_1 \ldots X_{m}). \label{aux94u}
\end{align}
where $\approx_{\delta}$ denotes that the two quantities differ by at most $\delta$. 

Let us now further assume that the mutual information $I(X_{m-1}:X_1):=S(X_{m-1})+S(X_1)-S(X_{m-1}X_1)$ of the disjoint regions $X_{m-1}X_1$ is small and upper bounded by $\veps$ (this assumption is justified by the finite correlation length of the state, but it is not necessary in the rigorous proof). Then
\begin{align} 
 &S(X_{m}|X_{m-1}) +  S(X_{1}|X_{m}) - S(X_1) \nonumber\\ 
 & \approx_{2\varepsilon}  S(X_{m-1} X_m X_1) - S(X_m). \label{aux66} 
\end{align}
Combining Eqs. (\ref{aux94u}) and (\ref{aux66}) we finally find
\begin{align}
S(\rho_X \Vert e^{-H_X}) &\approx_{m\varepsilon } I(X_2\ldots X_{m-2}:X_m|X_1X_{m-1}) \nonumber  
\\ &\approx 2\gamma.
\end{align}
Naively, this seems to finish proof of Eq.~\eqref{TEErelent}, however there are subtle problems (in addition to the extra assumption used that far away regions have small mutual information). First, the state $e^{-H_X}$ is not normalized. Second, there is no guarantee that there is no other choice of $H_X$ which significantly reduces the distance. 

The full proof is more involved and uses not merely strong subadditivity as in the previous argument, but also a recent strengthening of subaddiviity~\cite{Fawzi2015}. In the companion paper~\cite{KB16}, we prove the following. 
\blem\label{cmirelative}{(Theorem 3. in Ref.~\cite{KB16})}
Consider a 1D spin chain $X=X_1X_2...X_m$ with the size $N=|X_1...X_m|$. Let $\rho_{X_1...X_m}$ be a state such that 
\begin{align}
    I(X_{i+1}:X_{i+3}...X_{i-1}|X_{i+2})_\rho\leq\veps
\end{align}
for all $i\in[1, m]$.   
Define the set of Gibbs states of short-range Hamiltonians with interaction strength $K$ as 
\begin{equation*}
E^K_{nn}:=\left\{e^{-H}\left|\; H=\sum_ih_{X_iX_{i+1}},\; \|h_{X_iX_{i+1}}\|\leq K\right.\right\}\,.
\end{equation*}
Note that here we include the normalization factor in the Hamiltonian so that $\tr(e^{-H})=1$. 
Then, for $K=\Theta(N)$ and sufficiently small $\veps>0$, there exists a constant $c>0$ such that  for any tripartition $ABC$ of $X$ such that $B$ separates $A$ from $C$, it holds that
\begin{equation}\label{ap:thm4}
\min_{e^{-H}\in E^K}S\left(\rho_{X}\left\| e^{-H}\right.\right)=I(A:C|B)_\rho +\epsilon(N,\delta)
\end{equation}
and
\begin{equation*}
\left|\epsilon(N,\delta)\right|\leq cN^{\frac{5}{2}}\delta^{\frac{1}{16}}\,,
\end{equation*}   
where $\delta=8\sqrt{\veps}+2^{-N}$.
\elem
Our assumption on the area law~\eqref{arealaw} guarantees $\veps=\exp(-l/\xi)$ and $I(A:C|B)_\rho\approx2\gamma$. We can choose $l=\Th(\log N)$ so that $\epsilon(N,\delta)$ decays as in the order of $e^{-\Th(l)}$. Therefore, by applying Lemma~\ref{cmirelative} to our setting of Fig.~\ref{ring}, we obtain Eq.~\eqref{TEErelent}. 

 The idea of the proof of Lemma~\ref{cmirelative} is that explicitly constructing a state $\tpi_{X}\in E_{nn}^K$ such that $(i)$ has small $I(A:C|B)_\tpi$ and $(ii)$ locally indistinguishable from $\rho_X$. Suppose that such $\tpi_X$ exists. Then we can show that
\begin{align}
    I(A:C|B)_\rho&=S(AB)_\rho+S(BC)_\rho-S(B)_\rho-S(ABC)_\rho\\
    &\approx I(A:C|B)_\tpi +S(ABC)_\tpi-S(ABC)_\rho\\
    &\approx S(ABC)_\tpi-S(ABC)_\rho \label{ap:cmiformula}.
\end{align}
The first approximation follows from the Fannes inequality: two states which are close in the trace distance have almost same entropy. The second approximation follows from the assumption $(i)$. Let us next evaluate the LHS of Eq.~\eqref{ap:thm4}. For $\mu_X=e^{-\sum_{i}h_{X_iX_{i+1}}}\in E_{nn}^K$, it holds that
\begin{align}
    S(\rho\|\mu)&=-Tr[\rho_{AB}H_{AB}]-Tr[\rho_{BC}H_{BC}]-S(ABC)_\rho\\
    &\approx-Tr[\tpi_{AB}H_{AB}]-Tr[\tpi_{BC}H_{BC}]-S(ABC)_\rho\\
    &=S(\tpi\|\mu)+S(ABC)_\tpi-S(ABC)_\rho~\label{ap:dist}\,,
\end{align}
where $H_{AB}(H_{BC})$ is a sum of $h_{X_iX_{i+1}}$ within $AB(BC)$ so that $H=H_{AB}+H_{BC}$. 
The approximation follows since $\rho_{AB}(\rho_{BC})$ and $\tpi_{AB}(\tpi_{BC})$ are close in the trace distance. By taking minimum over $\mu_X\in E_{nn}^K$, the first term in Eq.~\eqref{ap:dist} vanishes since $\tpi_X\in E_{nn}^K$.  Therefore Eq.~\eqref{ap:cmiformula} and Eq.~\eqref{ap:dist} implies
\begin{align}
    \min_{\mu\in E_{nn}^K} S(\rho_X\|\mu)\approx S(ABC)_\tpi-S(ABC)_\rho \approx I(A:C|B)_\rho\,,
\end{align}
which completes the proof by $I(A:C|B)_\rho\approx2\gamma$ (more accurate evaluations of the errors are in Ref.~\cite{KB16}).

 The remaining problem is how to construct such $\tpi_X$. This part of the proof relies on the recently improved bound on the conditional mutual information:
\blem~\cite{Fawzi2015} \label{fawzirenner}
For any tripartite state $\rho_{ABC}$ in a finite-dimensional quantum system, 
\begin{equation} \label{eq:recovery} 
I(A:C|B)_{\rho}  \geq \min_{ \Delta : B \rightarrow BC} -2 \log F(\rho_{ABC}, \Delta_{B \rightarrow BC}(\rho_{AB})),
\end{equation}
where the minimum is over all completely-positive and trace-preserving (CPTP) map $\Delta_{B \to BC}:\cB({\cal H_B}) \to \cB({\cal H}_B \otimes {\cal H}_C)$, and $F(\rho, \sigma) := \tr((\sigma^{1/2} \rho \sigma^{1/2})^{1/2})$ is the fidelity. 
\elem  
This lemma means that a state with small conditional mutual information can be approximately generated from its reduced state by applying a CPTP map on the conditioning system only. Conversely, by using the Fannes inequality, the conditional mutual information is small if such a CPTP-map exists.  
By assumption~\eqref{arealaw}, we have 
\begin{align}
    I(A:B_2|B_1)_\rho&\leq e^{-\Th(l)}\\
    I(B_1:C|B_2)_\rho&\leq e^{-\Th(l)}\,.
\end{align}
Lemma~\ref{fawzirenner} reads there exists CPTP-maps $\Delta_{B_2\to B_2C}$ and $\Delta_{B_1\to AB_1}$ satisfying inequalities like Eq.~\eqref{eq:recovery}. By using these two CPTP-maps, we construct a state on $ABC$ by
\begin{align}
    {\tilde \rho}'_{ABC}:=\Delta_{B_2\to B_2C}\circ\Delta_{B_1\to AB_1}(\rho_B)\,.
\end{align}
 By construction, ${\tilde \rho}'_{ABC}$ and $\rho_{ABC}$ have similar reduced states on $AB$ and $BC$. Furthermore, one can show that ${\tilde \rho}_{ABC}'$ has small $I(A:C|B)$. This constructed state is not a Gibbs state in $E_{nn}^K$. We thus consider a Hamiltonian as in Eq.~\eqref{eq:entHamil} by using the reduced state of ${\tilde \rho}'_{ABC}$ and the corresponding (normalized) Gibbs state $\tpi_X$. It can be shown such $\tpi_X$ satisfies the required conditions~\cite{KB16}.
 
When TEE is strictly positive, $\min_{\mu\in E_{nn}^K}S\left(\rho_X\left\| \mu\right.\right)>0$ for any $N$ and therefore $H_{\rho_X}$ must contain non-local interactions. 
While we have not obtained a complete proof, we expect that the non-local part of $H_{\rho_X}$ is dominated by $m$-body interactions. 
To address this question, let us again set $A\equiv X_1$, $B\equiv X_2X_3X_{m-1}X_m$ and $C$ as the remaining subsystems. 
In a similar way to Eq.(3), we can show that 
\begin{align}\label{eq:abbc}
 \min_{\mu \in E^K_{AB,BC}}
 S(\rho_{X}\left\|\mu\right)\approx 2\gamma\,.
\end{align}
Here, $E^K_{AB,BC}$ is a set of Gibbs states of nearest-neighbor Hamiltonians $H_{AB}\ot I_C+I_A\ot H_{BC}$.  
Therefore, it contains at most $(m-1)$-body interaction acting on $BC=X_2X_3...X_m$. 
Eq.~\eqref{eq:abbc} implies that adding $(m-1)$-body interactions cannot improve the minimization in Eq.~\eqref{cmirelative}. This fact suggests
that the non-local part in the entanglement Hamiltonian is dominated by genuine $m$-body interactions.


\noindent
\section{Proof of Eq.~\eqref{thm1} and Eq.~\eqref{eq:thm2}}\label{ap2}

We give a proof of Eq.~\eqref{thm1} below. Eq.~\eqref{eq:thm2} can be proven in the exactly same way. 
\begin{proof}
Since $|\psi_{YXY\rq{}}\>$ is pure, it holds that
\begin{equation}
\lambda(\rho_{YY\rq{}})=\lambda(\rho_X)\,.
\end{equation}
As discussed in the main text, we have that
\begin{equation}
I(Y:Y\rq{})_\rho\leq e^{-\Th(l)}\,.
\end{equation}
The mutual information can be rewritten as 
\begin{equation}
I(Y:Y\rq{})_\rho= S(\rho_{YY'}\|\rho_Y\ot\rho_{Y'})\,.
\end{equation}
Therefore, we obtain 
\begin{equation}
\|\rho_{YY\rq{}}-\rho_Y^{\ot2}\|_1\leq e^{-\Th(l)}\,,
\end{equation}
by Pinsker inequality and the reflection symmetry which ensures $\rho_Y=\rho_{Y'}$.

For bounded Hermitian operators $A$ and $B$, the difference of their spectrum is bounded as
\begin{equation}\label{lemmapetz}
\left\|\lambda(A)-\lambda(B)\right\|_1\leq\|A-B\|_1\,
\end{equation}
(see e.g., Lemma 1.7 in Ref.~\cite{ohya2004quantum}).
Therefore, we obtain that
\begin{equation}
\left\|\lambda(\rho_{Y}^{\ot2})-\lambda(\rho_X)\right\|_1\leq e^{-\Th(l)}\,.
\end{equation}
Theorem~\ref{cmirelative} implies that $\rho_X$ is close to $e^{-H_X}/\tr e^{-H_X}$, where 
$H_X$ is short-ranged if $\gamma=0$ and otherwise contains non-local terms. 
By using Pinsker inequality and the triangle inequality, we obtain that 
\begin{align}
\left\|\lambda(\rho_Y^{\ot 2})-\lambda\left(e^{-H_X}\right)\right\|_1&\leq \left\|\rho_Y^{\ot 2}-\rho_X\right\|_1+\left\|\rho_X-e^{-H_X}\right\|_1\\
&\leq e^{-\Th(l)}\,. 
\end{align}
Let us introduce another cut-off to the spectrum which bounds from below
\begin{equation}
\lambda_\Lambda(A):=\left\{\lambda\in \lambda(A) \left| \lambda\geq \frac{1}{\Lambda}\right.\right\}\,.
\end{equation}
Clearly, 
\begin{equation}
\left\|\lambda_\Lambda(\rho_Y^{\ot 2})-\lambda_\Lambda\left(e^{-H_X}\right)\right\|_1\leq e^{-\Th(l)}\,. 
\end{equation}
Using the Lipschitz continuity of the logarithm in $[1/\Lambda,\infty)$, we conclude that
\begin{equation}
\left\|\lambda^\Lambda(-\log\rho_Y^{\ot 2})-\lambda^\Lambda\left(H_X\right)\right\|_1\leq \Lambda e^{-\Th(l)}\,. 
\end{equation}
Since $H^{(2)}_{\rho_Y}=-\log\rho_Y^{\ot 2}$, we complete the proof.
\end{proof}
\noindent
\section{The Entanglement Spectrum of MPS on A Cylinder}\label{ap3}
Here we argue that when a ground state  on a cylinder can be regarded as a MPS, the single entanglement spectrum of the half of a cylinder is equivalent to the spectrum of the boundary model. 

Let us consider a ground state on a cylinder with a boundary (or boundaries) as in the upper part of Fig.~\ref{cyl2}. 
As discussed in the main text, suppose that  $|\psi_{ N}\>$ can be written as a MPS:
\begin{align}\label{mps}
|\psi_{N}\>=\sum_{i_1,...,i_N}\left(L\right|A^{i_1}\ldots A^{i_N}\left|R\right)|i_1i_2\ldots i_N\>\,,
\end{align}
where the indices $\{i_j\}$ is associated with the  $j$th slice (column) of the cylinder, and 
$\{A^i\}_i$ are $D\times D$ matrices with a bond dimension $D\sim 2^r$. $|L)$ and $|R)$ are $D$-dimensional vector representing the boundary condition (we used ``)'' to distinguish them from vectors in physical systems).

Choose the first $m$ slices as subsystem $Y$. Then, one can show that in generic case the reduced density matrix on $Y$ of $|\psi_N\>$ and $|\psi_{\tilde N}\>$ are almost same for $N,\tilde N\gg1$.  
Therefore, the spectrum on $Y$ is approximately equivalent to the spectrum of the edge state defined for some fixed cylinder (Fig.~\ref{cyl2}). 
The reduced density matrix $\rho^{(N)}_{1\ldots m}=\tr_{m+1,...,N}|\psi_N\>\<\psi_N|$ on the first $m$ pieces of Eq.~\eqref{mps} is written as
\begin{align}
\rho^{(N)}_{1\ldots m}=&\sum_{i,j}(L|({\bar L}|\left(\prod_{k=1}^m(A^{i_k}\otimes{\bar A}^{j_k})\right){\mathbb T}^{N-m}|R)| {\bar R})\non\\
&\times|i_1i_2\ldots i_m\>\<j_1\ldots j_m|\,,
\end{align}
where  ${\mathbb T}:=\sum_i(A^i\otimes {\bar A}^i)\geq0$ is $D^2$-dimensional transfer matrix.

Let us estimate $\|\rho^{(N)}_{1\ldots m} - \rho^{({\tilde N})}_{1\ldots m}\|_1$ for ${\tilde N}\geq N$. For fixed $(i_1,j_1...,i_m,j_m)$, we have 
\begin{align}
&\left|(\rho^{(N)}_{1\ldots m} - \rho^{({\tilde N})}_{1\ldots m})_{(i_1,...,i_m)(j_1,...,j_m)}\right|\non\\
&=\left|(L|({\bar L}|\left(\prod_{k=1}^m(A^{i_k}\otimes{\bar A}^{j_k})\right)\left({\mathbb T}^{N-m}-{\mathbb T}^{{\tilde N}-m}\right)|R)| {\bar R})\right|\non\\
&\leq\left\|\left(\prod_{k=1}^m(A^{i_k}\otimes{\bar A}^{j_k})\right)\left({\mathbb T}^{N-m}-{\mathbb T}^{{\tilde N}-m}\right)\right\|_\infty\non\\
&\leq\prod_{k=1}^m\|A^{i_k}\|_\infty\|{\bar A}^{j_k}\|_\infty\left\|{\mathbb T}^{N-m}-{\mathbb T}^{{\tilde N}-m}\right\|_\infty\label{c1}
\end{align}
In the first inequality we used $\max_{i,j}|A_{i,j}|\leq\|A\|_\infty$ for the operator norm $\|\cdot\|$  (we assumed $(L|L)=(R|R)=1$).  
 We denote the eigenvalue decomposition of ${\mathbb T}$ by ${\mathbb T}=\sum_j\lambda_j|{\tilde j})({\tilde j}|$. 
For generic MPS, ${\mathbb T}$ has a unique maximal eigenvalue $\lambda_{\rm max}$ which we can set to be 1 without loss of generality. 
Then it holds that
\begin{equation}\label{c2}
\left\|{\mathbb T}^{N-m}-{\mathbb T}^{{\tilde N}-m}\right\|_\infty=\lambda_2^{\tilde N}-\lambda_2^N\leq\lambda_2^N\,,
\end{equation}
where $\lambda_2<1$ is the second largest eigenvalue of ${\mathbb T}$. By Inserting Eq.~\eqref{c2} to Eq.~\eqref{c1}, 
we have 
\begin{align}\label{c3}
&\left|(\rho^{(N)}_{1\ldots m} - \rho^{({\tilde N})}_{1\ldots m})_{(i_1,...,i_m)(j_1,...,j_m)}\right|\leq\cO(c^me^{-c'N})
\end{align}
for some nonnegative coefficients $c, c'$ determined by $A^i$ and $\lambda_2$. 
Remember that $\|A\|_1 \leq n^2\max_{i,j}|A_{ij}|$ for any $n\times n$ matrix $A$. $\rho^{(N)}_{1,...,m}-\rho^{(\tilde N)}_{1,...,m}$ is a $d^m\times d^m$ matrix. 
Therefore, we conclude from Eq.~\eqref{c3} that 
\begin{equation}\label{c4}
\|\rho^{(N)}_{1\ldots m} - \rho^{({\tilde N})}_{1\ldots m}\|_1\leq\cO\left((cd)^me^{-c'N}\right)\,.
\end{equation}
The upper bound is exponentially small with respect to $N-cm$ (up to a constant $c$).Note that the normalization factor for $\rho^{(N)}$ is given by 	
\begin{align}
\tr \rho^{(N)}&=(L{\bar L}|{\mathbb T}^N|R{\bar R})\\
&=(L{\bar L}|j_{1})(j_{1}|R{\bar R})\left(1+\cO\left(D^2e^{-c'N}\right)\right)\,.
\end{align}
Therefore the difference in Eq.~\eqref{c4} still holds after normalization.  

In summary, for any $N\geq m+l$, the entanglement spectrum on $Y$ of $|\psi_N\>$ is exponentially close to the spectrum of the edge state of $|\psi_{m+l}\>$ on $X=\{m+1,...,m+l\}$  with respect to $l$, the width of the edge $X$.  

\section{Exponentially Small Corrections in Area Law of Renyi-$\alpha$ Entropy}\label{ap4} 

In this appendix, we demonstrate the uniform area law of Renyi-$\alpha$ entropy (for every integer $\alpha \geq 2$) holds with exponentially small correction under an assumption which is expected to be true for generic 2D ground states in the topologically trivial phase. 
The main argument here is essentially one of the results in Ref.~\cite{PhysRevB.94.075151}, but we repeat it here for the completeness. 
For a state $\rho$, the Renyi-$\alpha$ entropy $S_\alpha(\rho)$ is defined by 
\begin{equation}
S_\alpha(\rho):=\frac{1}{1-\alpha}\log\tr(\rho^\alpha)\,. 
\end{equation}
Consider a translationally-invariant ground state $|\psi\>$ defined on a 2D lattice with size $N$. When a ground state is in the topologically-trivial phase, it can be (approximately) constructed from a product state only by a constant-depth local circuit~\cite{PhysRevB.72.045141,PhysRevB.82.155138}. 
In other words, there exists a set of unitaries $\{V_i\}$ (for each $N$) such that 
\begin{equation}\label{SRE}
|\psi\>=V_dV_{d-1}\ldots V_1|0\>^{\otimes N}\,,
\end{equation}
where $d$ is a constant of $N$ and each $V_i$ is 
\begin{equation}
V_i=\bigotimes_{k_i}V_i^{(k_i)}\,,
\end{equation}
a tensor product of local unitaries $V_i^{(k_i)}$ acting on disjoint sets of neighboring spins within radius $w=\cO(1)$. 

\begin{figure}[htbp]  
\begin{center}
\vspace{3mm}
\includegraphics[width=7cm]{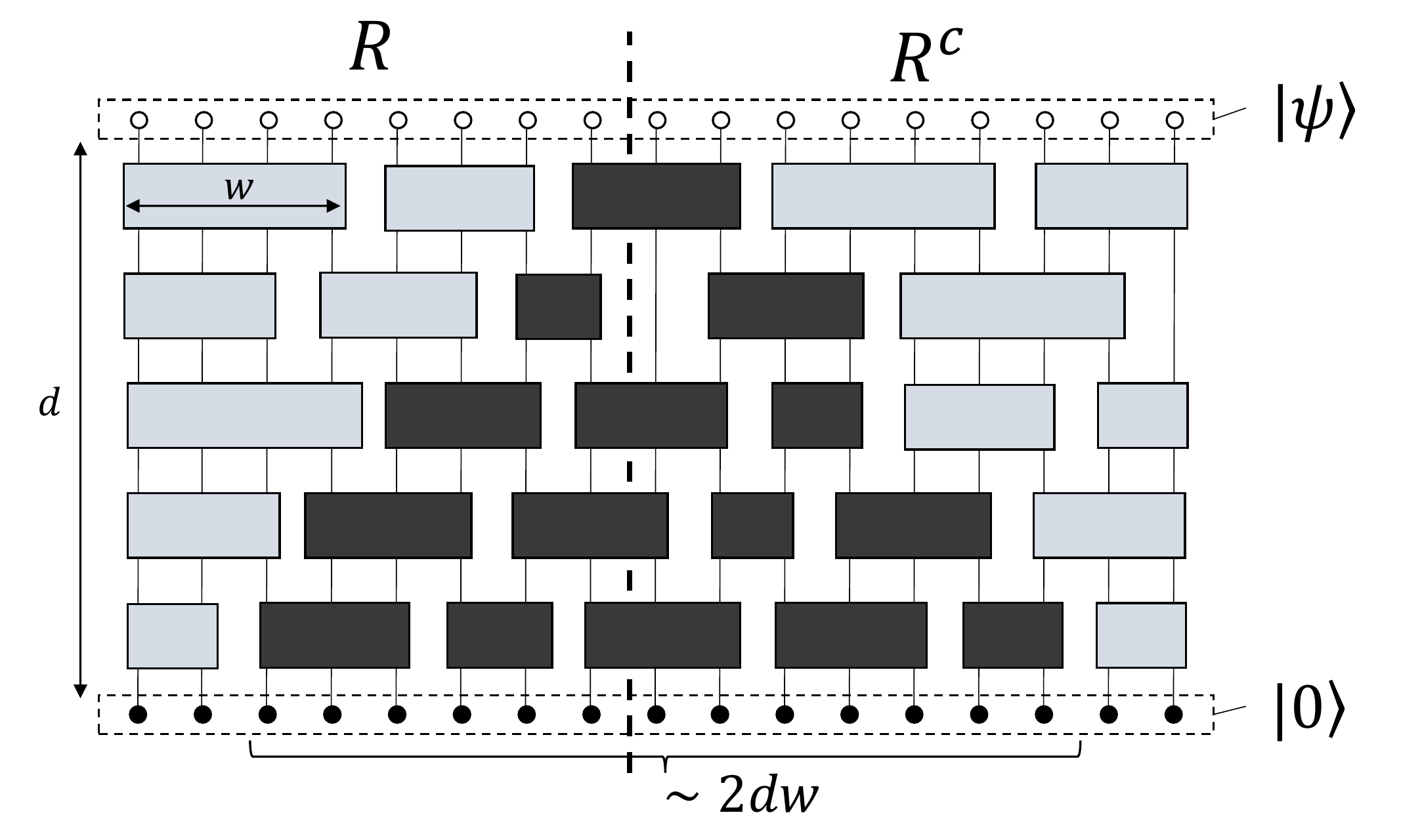}
\vspace{-5mm}
\end{center}
\caption{A schematic picture of reduction of the calculation of $S(R)_\rho$. The topologically trivial ground state $|\psi\>$ can be created by a product state $|0\>^{\ot N}$ by applying a constant-depth local circuit. The time step goes from bottom to top and each box represents a unitary matrix  $V_i^{(k_i)}$ acting on subsystems represented by vertical lines. When we divide systems into $R$ and $R^c$ (by the dotted line), only boxes colored by black contribute to the entanglement. We can remove all gray boxes ($U_{R}U_{R^c}$) without changing the entanglement entropy. Subsystems not acted by black boxes are then uncorrelated to all other systems. The state on the remained subsystems around the boundary is $|\phi_{RR^c}\>$. }
\label{mpsa}
\end{figure}

Let us divide the lattice into a square region $R$ (as in Fig.~1 in the main text) and its complement $R^c$ to calculate the entanglement entropy $S(R)_\rho$.  
Eq.~\eqref{SRE} is then rewritten as $|\psi_{RR^c}\>=U_{R}U_{R^c}U_B|0\>^{\otimes N}$ 
such that $U_R$ $(U_{R^c})$ only non-trivially act on $R (R^c)$ and $U_B$ acts on spins within distance $2dw$ from the boundary of $R$ (Fig.~\ref{mpsa}). 
Entanglement between $R$ and $R^c$ is invariant under $U^{-1}_{R}U^{-1}_{R^c}$ and therefore $S(R)_\rho$ is equivalent to that of $U^{-1}_{R}U^{-1}_{R^c}|\psi_{RR^c}\>$. 
$U^{-1}_{R}U^{-1}_{R^c}|\psi_{RR^c}\>$ is a product of a state $|\phi_{RR^c}\>$ around $\partial R$ and $|0\>$s far from $\partial R$, which are irrelevant for the entanglement (Fig.~\ref{mpsa}).  

\begin{figure}[htbp]  
\begin{center}
\vspace{-3mm}
\includegraphics[width=7.0cm]{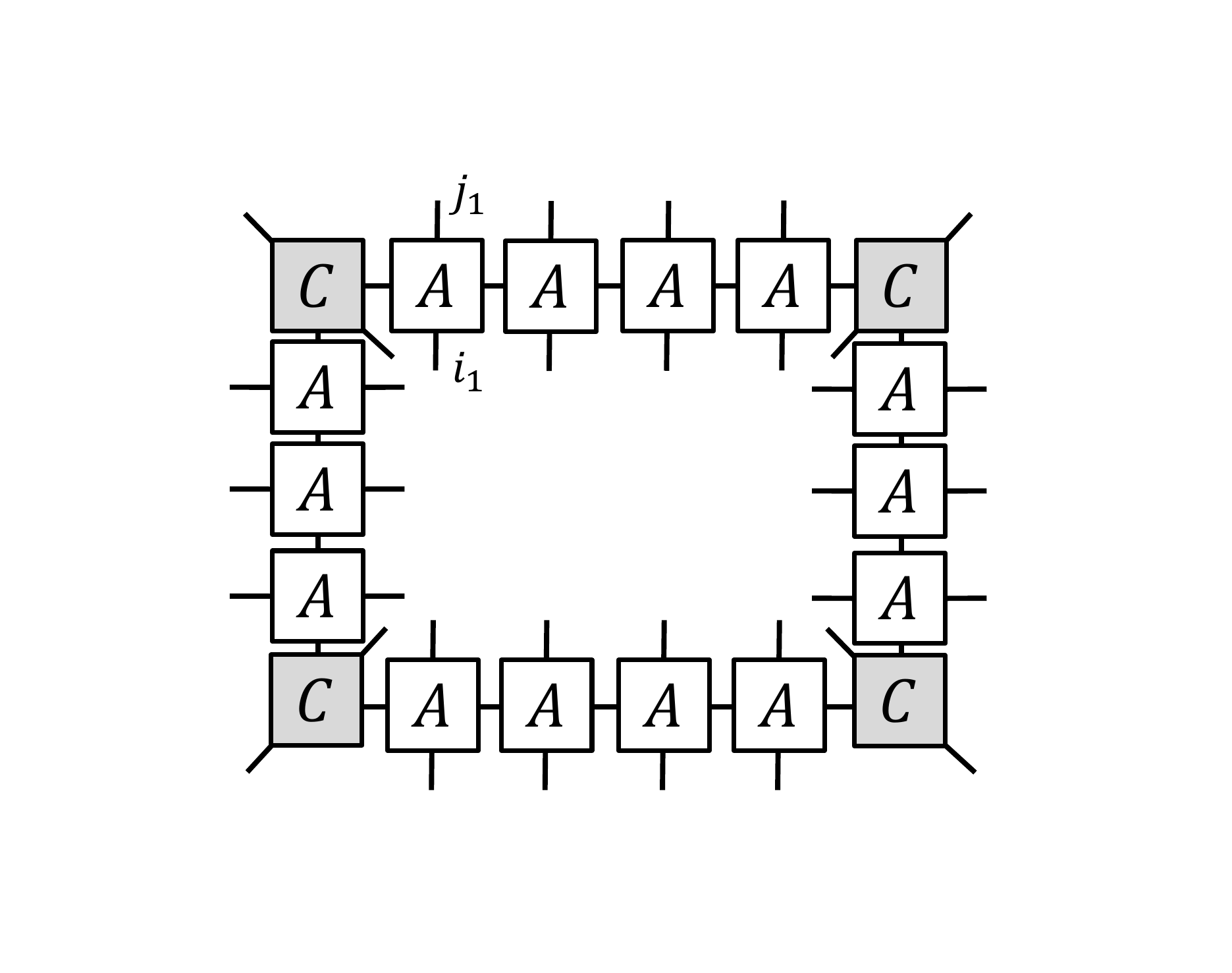}
\vspace{-5mm}
\end{center}
\caption{We can regard $\partial R$ as a periodic spin ladder under coarse-graining. $|\phi_{RR^c}\>$ is then represented as a MPS defined by two tensors $A$ and $C$ with a constant bond dimension. Each tensor has two legs corresponding either spins in $R$ or spins in  $R^c$. By tracing out the outer indices, we obtain a MPO representation of the reduced state $\phi_R$. }
\label{mpsb}
\end{figure}

From translationally-invariance, we expect $|\phi_{RR^c}\>$ can be written as a particular MPS: 
\begin{align}
|\phi_{RR^c}\>=\sum&\tr(A^{i_1j_1}\ldots A^{i_{l_1}j_{l_1}}C^{i_{l_1+1}j_{l_1+1}}\ldots C^{i_lj_l})\nonumber\\
&\times|i_1\ldots i_l\>_R|j_1\ldots j_l\>_{R^c}\,,
\end{align}
where tensor $A$ corresponds the edge and $C$ is associated the corner (Fig.~\ref{mpsb}).  By tracing out $R^c$ and taking $\alpha$-power, we obtain a matrix product operator (MPO) representation of $\phi^\alpha_R$. Its trace is given by 
\begin{equation}
\tr\phi^\alpha_R= \tr({\mathbb T}^{l_1}_\alpha{\mathbb T}^C_\alpha{\mathbb T}_\alpha^{l_2}{\mathbb T}_\alpha^C{\mathbb T}^{l_3}_\alpha{\mathbb T}^C_\alpha{\mathbb T}_\alpha^{l_4}{\mathbb T}^C_\alpha)\,,
\end{equation}
where ${\mathbb T}_\alpha:=\sum A^{i_1i_2}\otimes {\bar A}^{i_2i_3}\otimes \ldots \otimes {\bar A}^{i_{2\alpha-1} i_1}$ and ${\mathbb T^C}_\alpha$ is defined by replacing $A$ by $C$.  Generically, we expect that ${\mathbb T}_{\alpha}^{A}$ has an unique maximum eigenvalue $\lambda_{\max}(\alpha)$ yielding that
\begin{equation}
({\mathbb T}_\alpha^A)^l=\lambda_{\max}(\alpha)^l\left(|\lambda_{\max}(\alpha)\>\<\lambda_{\max}(\alpha)|+C+\cO(e^{-cl})\right)
\end{equation}
for a constant $c>0$. From this expression we can calculate the area law for Renyi-$\alpha$ entropy as
\begin{equation}
S_\alpha(R)_\rho=S_\alpha(R)_\phi=\frac{|\log\lambda_{\max}(\alpha)|}{\alpha-1}l+C+\cO(e^{-cl})\,,
\end{equation}
with $C$ a constant proportional to the number of corners of the region. 

This saturate the area law with a correction term which decays exponentially fast with respect to $l$ for fixed $\alpha$. 
Also, the coefficient of the linear term only depends on ${\mathbb T}$ and $\alpha$.  

The argument presented here does not apply to the von Neumann entropy, which is the case of relevance in our approach (since strong subadditivity only applies to it). But we believe that the correction $\veps$ in Eq.~(1) should hold also in that case, although a proof is left to future work.

\end{document}